\def\beg{\begin{equation}}
\def\eeq{\end{equation}}
\documentstyle[12pt]{article}
\begin{document}
\begin{center}
{\Large{\bf Comments on ``Mutually composite fermions in double layer quantum Hall systems", Jinwu Ye, cond-mat/0302558: Why it is wrong?}}
\vskip0.35cm
{\bf Keshav N. Shrivastava}
\vskip0.25cm
{\it School of Physics, University of Hyderabad,\\
Hyderabad  500046, India}
\end{center}

Jinwu Ye has shown that two flux quanta are attached in one layer while the electron is in the other layer to form a mutually composite fermion (MCF). This idea is based on an 
earlier idea that composite fermions (CF) are formed by attaching 
two flux quanta to one electron. We find that the
 formation of MCF is unphysical and it can not be the basis of a new 
theory. Similarly, the CF are also unphysical objects and their Lorentz
invariance is missing.
\vskip1.0cm
Corresponding author: keshav@mailaps.org\\
Fax: +91-40-2301 0145.Phone: 2301 0811.
\vskip1.0cm

\noindent {\bf 1.~ Introduction}

     Recently, Jinwu Ye[1] has suggested the idea of mutually composite fermions (MCF).  There are two layers of semiconductors, one layer has one electron and the other layer has two flux quanta. This object is called a mutually composite fermion (MCF). The electron to which two flux quanta are attached is called a composite fermion (CF).

     We show why both these ideas are wrong. Actually, earlier also we have pointed out as to why the CF is a wrong object.

\noindent{\bf 2.~~Description}

     The MCF is formed by keeping one electron in one layer and two
 flux quanta in another layer. One of the problems with this model 
is that the field is kept without any electric current. The flux, $\phi_o$ is related to the field and the area by the relation,
\beg
B.A=n\phi_o
\eeq
where $n$ is an integer. The flux quantum is equivalent to a field, $B=n\phi_o/A$, where $\phi_o=hc/e$ and ${\bf A}$ is the area. 
According to Biot and Savart's law of classical electrodynamics, an electron going in a circular orbit is equivalent to a current. The current of a charged particle is associated with mutually orthogonal electric and magnetic fields. Usually the cross product ${\bf E}\times{\bf H}$ gives the energy flow. So it is not possible to 
keep ${\bf B}$ in one layer and ${\bf E}$ in another. Jinwu Ye has suggested that ${\bf B}$ and ${\bf E}$ be kept in different layers and this is not correct. If it is correct, then how the ${\bf B}$ is 
carried to another layer and what will be its velocity? The velocity 
of ${\bf E}\times{\bf H}$ is the velocity of light but when ${\bf E}$ and ${\bf H}$ are decoupled, they will have different velocities. In order to generate a magnetic field,  a coil is made so that the current is along the coil and the magnetic field is along the axis of the coil but not at a distance such as $d$, where $d$ is the distance betweeen layers.

     Let us say that we have two flux quanta. These flux quanta have fields $B_1=n\phi_{o1}/A$ and $B_2=n\phi_{o2}/A$. The electric field associated with these flux quanta is zero, $E_1=E_2=0$. In that case ${\bf E}\times {\bf B}=0$ so that light can not come from such flux quanta. These flux quanta will be ``black" objects. According to one 
of the Maxwell equations $\nabla.{\bf H}=0$.  Therefore, 
$H_z=constant$.
If electron is kept in one layer and the flux quanta in another, then $H_z$=constant, condition is not satisfied so that $\nabla.{\bf H}=0$ condition is not obeyed.

     We have noted [2] several times that CF model is incorrect. In 
this model[3], two flux quanta are attached to one electron but there 
is no detachment of flux quanta from any where. Therefore, it can not conserve energy and extra flux quanta can not be created.

     It is perfectly justified to add the gradient of a function to the vector potential of electromagnetic field subject to the condition that a time derivative of the same function devided by the velocity of light is substracted from the scalar potential. In some of the works the latter condition is not met. They have set the correction to the scalar potential as zero so that the function becomes independent of time. 
Then only the vector potential or the magntic field has to be 
corrected. This is called the Chern-Simon theory but it does not 
satisfy the Lorentz condition.

     E-B decoupling. In the Maxwell equations, the time derivative of the magnetic field gives the space derivative of the electric field. Similarly, the time derivative of the electric field determines the gradient of the magnetic field and the current. Therefore, the electric field ${\bf E}$ determines ${\bf B}$ and the time derivative of ${\bf B}$ determines ${\bf E}$. This is the ${\bf E}$ and ${\bf B}$ coupling.
Therefore, it is not possible to put magnetic field or flux quanta at a distance from the current. The flux quanta are thus not separated so 
that  Jinwu Ye[1]  is not correct. This means that there is no theory 
of quantum Hall effect. Laughlin's incompressible wave function gives the  concept of a fractional charge but the quantity which was thought to be $e$ is $e/a_o^2$, where $Ba_o^2=n\phi_o$. So that the fraction 
may arise from ${\bf B}$. The only other theoretical model is due to Jain[3] which requires that two flux quanta are attached to one electron. This model is unphysical and can not form the basis of a new principle.

The proper explanation of the quantum Hall effect data is given in ref.4.

\noindent{\bf3.~~ Conclusions}.

     In conclusion, we find that there is no way for 
Jinwu Ye or Jain to be correct. They are badly inconsistent. There is no way to attach two flux quanta to one electron. Similarly, it is not possible to put flux in one layer and the electron in another. 
According to the classical electrodynamics, the ${\bf E}$ and
 ${\bf B}$ should be orthogonal and have one origin but not separated from each other.

\vskip1.25cm

\noindent{\bf4.~~References}
\begin{enumerate}
\item Jinwu Ye, cond-mat/0302558.
\item K. N. Shrivastava, cond-mat/0209666, 0210320, 0211223, 0310380, 0302009, 0302315; Bull. Am. Phys. Soc. J1.209(March 2003).
\item S. S. Mandal and J. K. Jain, Phys. Rev. B {\bf 66}, 155302(2002).
\item K.N. Shrivastava, Introduction to quantum Hall effect,\\ 
      Nova Science Pub. Inc., N. Y. (2002).
\end{enumerate}
\vskip0.1cm

Note: Ref.4 is available from:\\
 Nova Science Publishers, Inc.,\\
400 Oser Avenue, Suite 1600,\\
 Hauppauge, N. Y.. 11788-3619,\\
Tel.(631)-231-7269, Fax: (631)-231-8175,\\
 ISBN 1-59033-419-1 US$\$69$.\\
E-mail: novascience@Earthlink.net

\end{document}